\documentclass[english,reprint,aps]{revtex4-2}
\usepackage[T1]{fontenc}
\usepackage[utf8]{inputenc}
\usepackage[letterpaper]{geometry}
\usepackage{babel}
\usepackage{amsmath}
\usepackage{amssymb}
\usepackage{graphicx}
\usepackage[pdfusetitle,
 bookmarks=true,bookmarksnumbered=false,bookmarksopen=false,
 breaklinks=false,pdfborder={0 0 1},backref=false,colorlinks=false]
 {hyperref}

\makeatletter

\newcommand{\lyxmathsym}[1]{\ifmmode\begingroup\def\b@ld{bold}
  \text{\ifx\math@version\b@ld\bfseries\fi#1}\endgroup\else#1\fi}

\usepackage{babel}
\usepackage{xparse}

\newmuskip\pFqmuskip
\newcommand*{\pFq}[6][8]{%
	\begingroup 
	\pFqmuskip=#1mu\relax
	\mathchardef\normalcomma=\mathcode`,
	\mathcode`\,=\string"8000
	\begingroup\lccode`\~=`\,
	\lowercase{\endgroup\let~}\pFqcomma
	{}_{#2}F_{#3}{\left[\genfrac..{0pt}{}{#4}{#5};#6\right]}%
	\endgroup
}
\newcommand{\pFqcomma}{{\normalcomma}\mskip\pFqmuskip}

\ExplSyntaxOn
\NewDocumentCommand{\MeijerG}{smmmm}
{
	\IfBooleanTF{#1}
	{
		\vic_meijerg:nnnnnn { #2 } { #3 } { #4 } { #5 } { small } { }
	}
	{
		\vic_meijerg:nnnnnn { #2 } { #3 } { #4 } { #5 } { } { \; }
	}
}

\seq_new:N \l__vic_meijerg_args_in_seq
\seq_new:N \l__vic_meijerg_args_out_seq

\cs_new_protected:Nn \vic_meijerg:nnnnnn
{
	\seq_set_split:Nnn \l__vic_meijerg_args_in_seq { | } { #3 }
	\seq_clear:N \l__vic_meijerg_args_out_seq  
	\seq_map_inline:Nn \l__vic_meijerg_args_in_seq
	{
		\seq_put_right:Nn \l__vic_meijerg_args_out_seq
		{
			\begin{#5matrix} ##1 \end{#5matrix}
		}
	}
	G\sp{#1}\sb{#2}
	\left(
	\seq_use:Nn \l__vic_meijerg_args_out_seq { #6\middle|#6 }
	#6\middle|#6
	#4
	\right)
}
\ExplSyntaxOff

\makeatother

\begin{document}
\title{Tunable nonlinear excitonic optical response in biased bilayer graphene}
\author{M. F. C. Martins Quintela$^{1,2,3}$}
\email{mfcmquintela@gmail.com}
\author{N. M. R. Peres$^{1,2,4}$
}
\author{T. Garm Pedersen$^{3}$
 }
\address{$^{1}$Department of Physics and Physics Center of Minho and Porto
Universities (CF--UM--UP), Campus of Gualtar, 4710-057, Braga, Portugal}
\address{$^{2}$International Iberian Nanotechnology Laboratory (INL), Av. Mestre
Jos{é} Veiga, 4715-330, Braga, Portugal}
\affiliation{$^{3}$Department of Materials and Production, Aalborg University,
9220 Aalborg {Ø}st, Denmark}
\affiliation{$^{4}$POLIMA--Center for Polariton--driven Light--Matter Interactions, University of Southern Denmark, Campusvej 55, DK-5230 Odense M, Denmark}
\begin{abstract}
Biased bilayer graphene (BBG) is an important system for studies of excitonic effects in graphene--based systems, with its easily tunable bandgap. This bandgap is governed by an external gate voltage, allowing one to tune the optical response of the system. In this paper, we study the excitonic linear and nonlinear optical response of Bernal stacked BBG as a function of the gate voltage, both for in--plane (IP) and out--of--plane (OOP) directions. Based on a semi-analytical model of the electronic structure of BBG describing the influence of gate voltage on excitonic binding energies, we focus our discussion on both the IP and OOP excitonic response. Both linear and second harmonic generation (SHG) nonlinear responses are shown to be very sensitive to the gate voltage, as both the interband momentum matrix elements and the bandgap of the system will vary greatly with bias potential. 
\end{abstract}
\maketitle

\section{Introduction}

Since the first mechanical isolation of graphene\cite{novoselov_electric_2004}, many different two--dimensional (2D) materials have been studied\cite{doi:10.1021/acsnano.1c00344}, such as hexagonal Boron Nitride (hBN)\cite{caldwell_photonics_2019} and transition metal dichalcogenides (TMDs)\cite{RevModPhys.90.021001}. This interest then recently transitioned into layered materials with broken vertical symmetry, which have been considered both from theoretical and experimental perspectives. These materials include, but are not limited to, buckled monolayers \cite{siahin_monolayer_2009,PhysRevB.99.085432,le_fracture_2021,kezerashvili_effects_2021}, Janus materials \cite{Lu2017,Zhang2017,doi:10.1021/acs.nanolett.0c03412,doi.org/10.1002/lpor.202100726,Shi2023}, hetero- and biased homobilayers \cite{doi:10.1021/nl902932k,Gong2014,Rivera2016,PhysRevB.91.205405}.
In stark contrast to its monolayer counterpart, BBG has a tunable gap which can be as large as $150\,\mathrm{meV}$\cite{park_tunable_2010}. The magnitude of this bandgap can be controlled via an external electric field and dielectric environment, and allows the formation of tunable excitons, which have already been both measured experimentally\cite{ju_tunable_2017} and described theoretically\cite{doi:10.1021/nl902932k,Avetisyan2018,henriques_absorption_2022,sauer_exciton_2022,PhysRevB.107.085104}.

As BBG presents an intrinsic OOP asymmetry due to the interlayer bias potential, one can directly study the effects of this asymmetry on the excitonic response. Tuning the interlayer bias potential, one can alter both the electronic structure of BBG and the interlayer asymmetry, probing the effects on both IP and OOP optical response\cite{doi:10.1143/JPSJ.78.104716}. 
The probing of the OOP response, namely the OOP non--linearities, leads to additional degrees of freedom useful for vertical photonics structures\cite{Kleemann2017,doi:10.1021/acs.nanolett.8b02652}, allowing for novel approaches in the design of ultrafast optical devices\cite{doi.org/10.1002/lpor.202100726}, such as non--linear holograms\cite{doi:10.1021/acs.nanolett.9b02740}, broadband ultrafast frequency converters\cite{Soavi2018,Klimmer2021}, miniaturized logic gates\cite{doi:10.1126/sciadv.abq8246,Li2022}, among others. Furthermore, the presence of various non--zero tensor components which include non--diagonal components also provides a greater freedom when experimentally probing the optical response of BBG.

Recent theoretical works\cite{PhysRevB.108.235401} have reviewed the effects of gate voltage on the free--carrier injection rate and shift current in AA- and AB- (Bernal) stacked bilayer graphene, with many other theoretical works focusing on single--particle properties of bilayer graphene structures\cite{doi:10.1126/science.1130681,PhysRevB.73.245426,PhysRevB.78.085432,ROZHKOV20161,https://doi.org/10.1002/qute.202100056,PhysRevLett.99.216802}, especially when an interlayer twist angle is introduced\cite{PhysRevLett.99.256802,Cao2018,MartinsQuintela2020,Andrei2020,PhysRevB.106.035401}. However, and as in many layered materials, the optical response of BBG is dominated by excitons\cite{yu_valley_2015} and their effects are, therefore, fundamental in the study of the optical response of this material. In this paper, we focus on the effects of gate voltage on the excitonic optical response of AB-stacked bilayer graphene, considering both linear and SHG nonlinear response. 

This paper is organized as follows: we begin, in Section \ref{sec:hamiltonian}, by discussing the electronic structure and its dependence on the external gate voltage. This is then followed by the computation of the momentum and Berry connection matrix elements, both for IP and OOP directions. Knowing the momentum matrix elements, we compute the free--carrier linear response for both the full Hamiltonian and a reduced two--band model. This allows us to identify the  frequency ranges where two bands are enough to describe the optical response of the system as a function of the gate voltage. Afterwards, in Section \ref{sec:BSE}, we proceed to the computation of the excitonic states through solving the Bethe--Salpeter equation. This involves an approximation of the effective screening length in the electron--hole electrostatic potential, followed by a discussion on the distinction between intralayer and interlayer phenomena. Finally, in Section \ref{sec:optical}, we compute the linear and SHG nonlinear excitonic optical response of BBG. After obtaining the IP and OOP optical selection rules for both linear and nonlinear response, we discuss the effects of gate voltage, as well as the relative amplitudes of the different elements of the nonlinear conductivity tensor. 

\section{Single Particle Biased Bilayer Graphene Hamiltonian \label{sec:hamiltonian}}

In the basis $\left\{ \left|1,b\right\rangle ,\left|2,b\right\rangle ,\left|1,t\right\rangle ,\left|2,t\right\rangle \right\} $,
where $b/t$ denotes the bottom/top layer and $1/2$ denotes the different
sublattices in each layer, the Hamiltonian of Bernal stacked BBG\cite{McCann_2013} in the Dirac approximation around the Dirac points $\left(K_x,\tau K_y\right)$, with $\tau=\pm1$ the valley index, is given by
\begin{equation}
\mathcal{H}=\left(\begin{array}{cccc}
V & -\gamma_{0}f & 0 & -i\gamma_{3}f^{*}\\
-\gamma_{0}f^{*} & V & \gamma_{z} & 0\\
0 & \gamma_{z} & -V & -\gamma_{0}f\\
i\gamma_{3}f & 0 & -\gamma_{0}f^{*} & -V
\end{array}\right),\label{eq:single_ham}
\end{equation}
where $V$ is the bias potential applied to the bilayer system, $\gamma_{0}=3.12\,\mathrm{eV}$
is the intralayer nearest--neighbour hopping parameter, and the interlayer
parameters $\gamma_{z}=0.377\,\mathrm{eV}$ (interlayer direct vertical
coupling) and $\gamma_{3}=0.377\,\mathrm{eV}$ (skew coupling parameter). Restricting our analysis to the Dirac cones $K,K^{\prime}$, we consider $\mathbf{k}$ measured from the Dirac points and write
\begin{align*}
	f\left(\mathbf{k}\right) & \approx\frac{\sqrt{3}}{2}a\left(k_{x}-i\tau k_{y}\right)
\end{align*}
with $\hbar v_{F}=\frac{\sqrt{3}}{2}a\gamma_{0}$
the Fermi velocity, the polar angle of $\mathbf{k}$ defined as $\theta=\tan^{-1}\left(\frac{k_{y}}{k_{x}}\right)$, and $a=2.46\,\text{\AA}$ the lattice parameter. The interlayer
bias potential $2V$ corresponds to placing the system in an electric
field $\frac{2V}{ed}$, with $d$ the interlayer separation. In this paper we will mainly consider $V=55\,\mathrm{meV}$ as it is at this bias that the dominant peak in the IP linear response matches with recent experimental results\cite{ju_tunable_2017}.

\begin{figure*}
	\centering \hspace{-0.3cm}\includegraphics[scale=0.72]{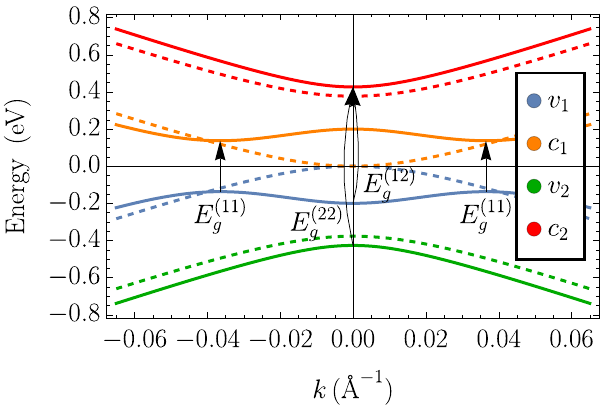} \quad\vspace{-0.1cm}\includegraphics[scale=0.72]{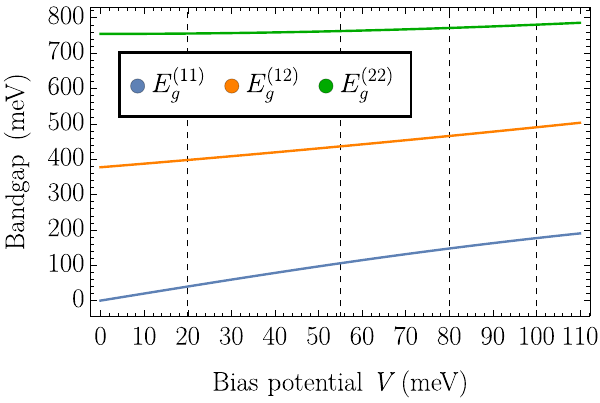}
	\caption{(Left) Band structure of BBG with a large bias $(V=200\,\mathrm{meV})$
		near the Dirac points. Coloured dashed lines correspond to the band
		structure under zero external bias, vertical arrows represent
		the three direct bandgaps. 
		(Right) Evolution of the three distinct direct bandgaps in BBG with increasing external bias near the Dirac points. Vertical dashed lines represent the four different values of $V$ applied in numerical studies. }\label{fig:band_structure}
\end{figure*}

The band structure under the Dirac approximation is given by 
\begin{align}
	E_{\lambda_\eta}&=\lambda\sqrt{\hbar^2 v_F^2 k^2+V^2+\frac{1}{2} \gamma_z^2+\left(2\eta - 3\right) \Lambda}\nonumber\\
	\Lambda&=\sqrt{\frac{1}{4} \gamma_z^4+\hbar^2 v_F^2 k^2\left(4 V^2+\gamma_z^2\right)},\label{eq:bandstructure}
\end{align}
where $\lambda=\pm1$ represents the conduction and valence bands ($\lambda=+1$ will be conduction (c) bands, $\lambda=-1$ valence (v) bands), and $\eta=1,2$ identifies the band closest ($\eta=1$) and furthest away ($\eta=2$) from the bandgap. Direct substitution of  $\eta=1,2$ into Eq. (\ref{eq:bandstructure}) leads to a factor of $\pm 1$, as expected for the band structure of this system. This band structure is presented in Fig. \ref{fig:band_structure}. Focusing on direct bandgaps, we write the generic bandgap as
\begin{equation}
	E_{g}^{(\eta,\eta^\prime)}=E_{c_\eta}-E_{v_{\eta^\prime}}.
\end{equation} 
Due to the nature of the BBG band structure, the minimum bandgap between the two $\eta=1$ bands will not occur at $k=0$ (see left panel of Fig. \ref{fig:band_structure}), instead occurring at 
\begin{equation}
k=\sqrt{2}\frac{V}{\hbar v_F}\sqrt{\frac{2V^{2}+\gamma_{z}^{2}}{4V^{2}+\gamma_{z}^{2}}}
\end{equation}
measured from the Dirac points. 
For instance, 
\begin{align}
	E_{g}^{(11)}&=\frac{2V\gamma_{z}}{\sqrt{4V^{2}+\gamma_{z}^{2}}}\approx106\,\mathrm{meV}\label{eq:bandgap}
\end{align}
for the considered bias potential $V=55\,\mathrm{meV}$.
The two larger direct bandgaps, namely $E_{g}^{(12)}$ and $E_{g}^{(22)}$,
will occur at $k=0$ and are significantly larger. For $V=55\,\mathrm{meV}$, these take the values 
\begin{align}
	E_{g}^{(12)}&=V+\sqrt{V^{2}+\gamma_{z}^{2}}\approx436\,\mathrm{meV},\nonumber\\
	E_{g}^{(22)}&=2\sqrt{V^{2}+\gamma_{z}^{2}}\approx762\,\mathrm{meV},
\end{align}
and $E_{g}^{(21)}=E_{g}^{(12)}.$
The three distinct bandgaps are plotted in the right panel of Fig. \ref{fig:band_structure} for external bias between $0$ and $110\,\mathrm{meV}$.
The large difference between the three bandgaps suggests that, as we will see ahead, the two $\eta=1$ bands will dominate the low--energy response of the system. 

\subsection{In--Plane Matrix Elements}

For simplicity, we will discard the contributions from skew coupling
to the eigenvectors, considering it instead as a perturbation when
computing dipole matrix elements. 
In a compact notation, we define the normalized eigenvectors for future
computations as 
\begin{equation}
\left|\lambda_{\eta}\left(\mathbf{k}\right)\right\rangle =
\left[\begin{array}{c}
e^{-2i\tau\theta}\alpha_{1,\lambda_\eta}\left(k\right)\\
e^{-i\tau\theta}\alpha_{2,\lambda_\eta}\left(k\right)\\
e^{-i\tau\theta}\alpha_{3,\lambda_\eta}\left(k\right)\\
\alpha_{4,\lambda_\eta}\left(k\right)
\end{array}\right].\label{eq:eigenvectors_simp}
\end{equation}
This gauge choice will carry a pseudo--spin\cite{park_tunable_2010,henriques_absorption_2022,di_sabatino_optical_2020,zhang_optical_2018,cao_unifying_2018,henriques_excitonic_2022}
factor of $m_{s}=-2$ in the $\tau=1$ valley. For compactness in the expressions for the momentum matrix elements and Berry connections, the $k$ dependence in the eigenvector components will be suppressed.

To compute the IP interband momentum matrix elements, chosen without
loss of generality in the $x$ direction, we must analyze 
\begin{equation}
	P_{v_{\eta}c_{\eta^{\prime}}\mathbf{k}}^{x}=\frac{m_{0}}{\hbar}\left<v_{\eta}\middle|\frac{\partial\mathcal{H}}{\partial k_{x}}\middle|c_{\eta^{\prime}}\right>,\label{eq:P_def}
\end{equation}
where $m_0$ is the free electron mass. 
The operator $\frac{\partial\mathcal{H}}{\partial k_{x}}$ can be readily
computed and is given by 
\begin{align*}
\frac{\partial\mathcal{H}}{\partial k_{x}} & =\left(\begin{array}{cccc}
0 & -\hbar v_{F} & 0 & -i\frac{\gamma_{3}}{\gamma_{0}}\hbar v_{F}\\
-\hbar v_{F} & 0 & 0 & 0\\
0 & 0 & 0 & -\hbar v_{F}\\
i\frac{\gamma_{3}}{\gamma_{0}}\hbar v_{F} & 0 & -\hbar v_{F} & 0
\end{array}\right).
\end{align*}
Yet, obtaining the matrix elements taking into account skew coupling
due to finite $\gamma_{3}$ in the eigenvectors becomes
extremely unwieldy. As a consequence, we consider an approximation where $\gamma_{3}$
is taken as a small perturbation to the Dirac BBG system and, therefore, its only effect will be introducing
additional terms in the momentum matrix elements proportional to $\frac{\gamma_{3}}{\gamma_{0}}\hbar v_{F}$, as presented in
$\frac{\partial\mathcal{H}}{\partial k_{x}}$. This follows the approach previously outlined in a similar study of the linear response of BBG\cite{henriques_absorption_2022}, where experimental results\cite{ju_tunable_2017} where accurately predicted under this same perturbative approach. This also means that skew coupling will play no part in the calculation of Berry connections, as it is only dependent on the eigenvectors defined in Eq. (\ref{eq:eigenvectors_simp}).

Focusing only on the $\eta=1$ bands, the exciton momentum matrix element
will be given explicitly as 
\begin{align}
	&P_{v_{\eta}c_{\eta^\prime}\mathbf{k}}^{x} =m_{0}v_{F}\left[e^{i\tau\theta}\left(\alpha_{1,v_\eta}^*\alpha_{2,c_{\eta^{\prime}}}+\alpha_{3,v_\eta}^*\alpha_{4,c_{\eta^{\prime}}}\right)\right.\nonumber \\
	&\,+e^{-i\tau\theta}\left(\alpha_{2,v_\eta}^*\alpha_{1,c_{\eta^{\prime}}}+\alpha_{4,v_\eta}^*\alpha_{3,c_{\eta^{\prime}}}\right)\label{eq:expansion_P}\\
	& \,\left.-i\frac{\gamma_{3}}{\gamma_{0}}\left(e^{2i\tau\theta}\alpha_{1,v_\eta}^*\alpha_{4,c_{\eta^{\prime}}}-e^{-2i\tau\theta}\alpha_{4,v_\eta}^*\alpha_{1,c_{\eta^{\prime}}}\right)\right].\nonumber
\end{align}
In this form, it is also clear from the phase factors that, upon multiplication by the conjugate and angular integration, the lowest contribution
from skew coupling to the linear response will be quadratic in $\left|\gamma_{3}/\gamma_{0}\right|\approx0.1$. As $\left|\gamma_{3}/\gamma_{0}\right|\ll1$, we will keep only the dominant term in each expansion. The validity of this perturbative approach has also been checked against a full tight--binding numerical calculation.

Comparing both the single--particle Hamiltonian as in Eq. (\ref{eq:single_ham}) and the momentum matrix element as given in Eq. (\ref{eq:expansion_P}) against those used in Ref. \cite{henriques_absorption_2022}, one can see that Ref. \cite{henriques_absorption_2022} is missing the $i$ factor in the skew coupling contribution for both the Hamiltonian and the momentum matrix element. Although the lack of this term is of no consequence for the linear response discussed in Ref. \cite{henriques_absorption_2022} due to the proportionality to $\left|\gamma_{3}/\gamma_{0}\right|^2$, this factor is of importance for the IP nonlinear response as it will be clear when the excitonic matrix elements are discussed. 

A different contribution, namely trigonal warping, consists of considering the quadratic term in the series expansion of $f$, leading to\cite{taghizadeh_nonlinear_2019} 
\begin{equation}
	f\left(\mathbf{k}\right)\approx\frac{\sqrt{3}a}{2}\left[\left(k_{x}-i\tau k_{y}\right)+i\zeta_{\mathrm{TW}}a\left(k_{x}-i\tau k_{y}\right)^{2}\right],\label{eq:trig_warp_f}
\end{equation}
where $\zeta_{\mathrm{TW}}=\frac{\sqrt{3}}{12}$ is a fixed numerical factor resulting from the series expansion of $f$.
Careful analysis of the effects of this quadratic term on the momentum matrix elements via Eq. (\ref{eq:P_def}) reveals that the phase factor associated will be $e^{\pm2i\tau\theta}$, identical to the contribution from skew coupling in Eq. (\ref{eq:expansion_P}). Furthermore, upon computation of the excitonic matrix elements, this quadratic correction proves to be negligible when compared against that originating from skew coupling and, therefore, it will be ignored. 

In the $k_{x}$ direction, the matrix elements of the Berry connection
between bands $n_\eta$ and $m_{\eta^\prime}$ read 
\begin{align}
\Omega_{n_{\eta}m_{\eta\prime}\mathbf{k}}^{x} & =i\left<n_{\eta}\middle|\frac{\partial}{\partial k_{x}}\middle|m_{\eta\prime}\right>.\label{eq:berry_generic}
\end{align}
The exact form of the Berry connection
will be needed when computing generalized derivatives\cite{aversa_nonlinear_1995} and $\Omega_{c_{\eta}c_{\eta^{\prime}}\mathbf{k}}^{x}$ will be given in terms of the eigenvector components as
\begin{widetext}
	\begin{align}
		\Omega_{c_{\eta}c_{\eta^{\prime}}\mathbf{k}}^{x}	
		&=i\cos\theta\left(\alpha_{1,c_\eta}^*\frac{\partial\alpha_{1,c_{\eta^{\prime}}}}{\partial k}+\alpha_{2,c_\eta}^*\frac{\partial\alpha_{3,c_{\eta^{\prime}}}}{\partial k}+\alpha_{3,c_\eta}^*\frac{\partial\alpha_{3,c_{\eta^{\prime}}}}{\partial k}+\alpha_{4,c_\eta}^*\frac{\partial\alpha_{4,c_{\eta^{\prime}}}}{\partial k}\right)\nonumber\\
		&	\quad-\frac{\tau}{k}\sin\theta\left(2\alpha_{1,c_\eta}^*\alpha_{1,c_{\eta^{\prime}}}+\alpha_{2,c_\eta}^*\alpha_{1,c_{\eta^{\prime}}}+\alpha_{3,c_\eta}^*\alpha_{3,c_{\eta^{\prime}}}\right),\label{eq:omega_cc_simp}
	\end{align}
\end{widetext}
and $\Omega_{v_{\eta}v_{\eta^{\prime}}\mathbf{k}}^{x}$ will be defined analogously for the valence bands.

\subsection{Out--of--Plane Matrix Elements}

Before computing the single particle linear optical response, we will discuss the OOP momentum and Berry connection. To this end, we begin by defining the matrix elements of $z$ from the extension of $\frac{d}{2}\sigma_{z}$\cite{PhysRevB.107.235416} to a bilayer system, with $\sigma_z$ the diagonal Pauli matrix, as 
\begin{align}
	z_{n_{\eta}m_{\eta\prime}\mathbf{k}} & =\frac{1}{2}\left\langle n_{\eta}\left|\mathrm{diag}\left[d,d,-d,-d\right]\right|m_{\eta^\prime}\right\rangle.
\end{align}
The interlayer separation $d$ has been previously discussed in the literature\cite{doi:10.1021/ja994457o,Alam_2011,Butz2014,PhysRevB.104.L180202},
with reported values for $d\approx3.46\,\text{\AA}$. The momentum matrix element then reads 
\begin{align}
	P_{v_{\eta}c_{\eta^{\prime}}\mathbf{k}}^{z} & =\frac{m_0}{i\hbar}E_{c_{\eta^\prime}v_{\eta}\mathbf{k}}z_{v_{\eta}c_{\eta^{\prime}}\mathbf{k}}.
\end{align}
Explicitly computing $z_{v_{\eta}c_{\eta^{\prime}}\mathbf{k}}$, we obtain 
\begin{align}
	P_{v_{\eta}c_{\eta^{\prime}}\mathbf{k}}^{z} & =\frac{m_0 d}{2i\hbar}E_{c_{\eta^\prime}v_{\eta}\mathbf{k}}\left[\alpha_{1,v_\eta}^*\alpha_{1,c_{\eta^{\prime}}}+\alpha_{2,v_\eta}^*\alpha_{2,c_{\eta^{\prime}}}\right.\nonumber \\ &\quad\left.-\alpha_{3,v_\eta}^*\alpha_{3,c_{\eta^{\prime}}}-\alpha_{4,v_\eta}^*\alpha_{4,c_{\eta^{\prime}}}\right].\label{eq:expansion_P_z}
\end{align}

For the nonlinear response, we must also compute the matrix elements
$z_{c_{\eta}c_{\eta^{\prime}}\mathbf{k}}$, $z_{v_{\eta}v_{\eta^{\prime}}\mathbf{k}}$. Similarly to $z_{v_{\eta}c_{\eta^{\prime}}\mathbf{k}}$, we write these in terms of the radial components of the eigenvectors as
\begin{align*}
	z_{c_{\eta}c_{\eta^{\prime}}\mathbf{k}} & =\frac{d}{2}\left[\alpha_{1,c_\eta}^*\alpha_{1,c_{\eta^{\prime}}}+\alpha_{2,c_\eta}^*\alpha_{2,c_{\eta^{\prime}}}\right.\\
	& \quad\left.-\alpha_{3,c_\eta}^*\alpha_{3,c_{\eta^{\prime}}}-\alpha_{4,c_\eta}^*\alpha_{4,c_{\eta^{\prime}}}\right],
\end{align*}
and analogous for $z_{v_{\eta}v_{\eta^{\prime}}}$. Hence, we obtain the Berry connection as\cite{aversa_nonlinear_1995} 
\begin{align}
	&\Omega_{c_{\eta}c_{\eta^{\prime}} \mathbf{k}}^{z}-\Omega_{v_{\eta}v_{\eta^{\prime}} \mathbf{k}}^{z} =z_{c_{\eta}c_{\eta^{\prime}}\mathbf{k}}-z_{v_{\eta}v_{\eta^{\prime}}\mathbf{k}}.\label{eq:berry_z}
\end{align}

\subsection{Single Particle Linear Optical Response}

In a clean semiconductor at zero temperature, the diagonal optical response
reads \cite{genkin1968contribution,aversa_nonlinear_1995,kirtman_extension_2000,margulis_optical_2013,pedersen_intraband_2015}
\begin{equation}
\sigma_{\beta\beta}\left(\omega\right)=\frac{e^{2}\hbar^{2}\omega}{i\pi^{2}m_{0}^{2}}\sum_{c,v}\int\frac{\left|P_{cv\mathbf{k}}^{\beta}\right|^{2}}{E_{cv\mathbf{k}}\left(E_{cv\mathbf{k}}^{2}-\hbar^{2}\omega^{2}\right)}d^{2}\mathbf{k},
\end{equation}
where $\sum_{c,v}$ sums over all conduction and valence bands. Within the two--band approximation, the sum over valence and conduction bands is dropped.
With the expansion near the Dirac cones, integration will now be
over the infinite Dirac cones and a sum over valleys must also be made.

\begin{figure}
	\centering\hspace{-0.864cm}\includegraphics[scale=0.7296]{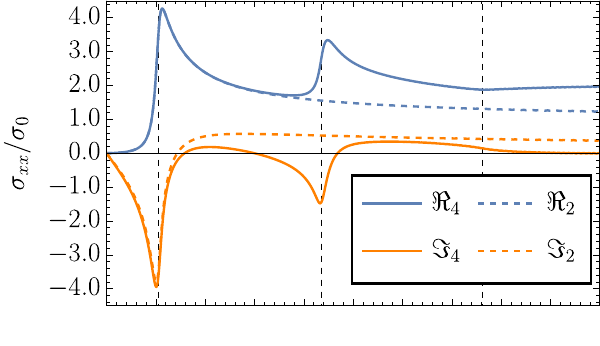}
	
	\vspace{-.655cm}\hspace{-0.599cm}\includegraphics[scale=0.7699]{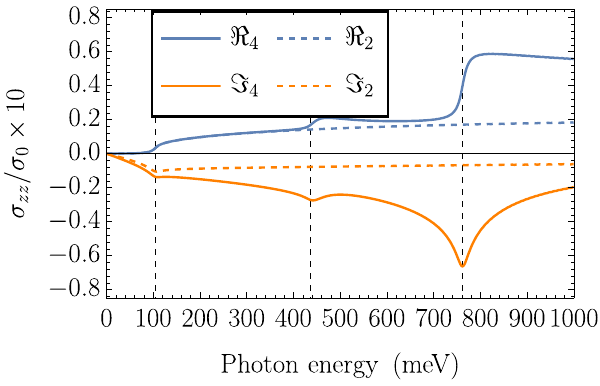}
	
	\caption{Comparison between full four--band calculation and calculation with
		only the two lowest bands of both IP (top) and OOP (bottom) free--carrier linear optical response for biased bilayer graphene with
		a bias potential $V=55\,\mathrm{meV}$ and broadening $\hbar\Gamma=5\,\mathrm{meV}$. Real part represented in blue, denoted $\Re_n$, and imaginary part in orange, denoted $\Im_n$, for the $n-$band calculation.  
		Vertical dashed lines represent the various bandgaps of the system, from left to right $E_{g}^{(11)}$, $E_{g}^{(12)}$, $E_{g}^{(22)}$.}\label{fig:single-part-cond}
\end{figure}

Inspecting Eq. (\ref{eq:expansion_P}), one sees that the angular integration is trivial. Although the radial integration proves much more complicated, one can easily perform it numerically and, 
introducing a broadening parameter via the transformation $\hbar\omega\rightarrow\hbar\omega+i\hbar\Gamma$ with $\hbar\Gamma=5\,\mathrm{meV}$, obtain both the free--carrier IP (top) and OOP (bottom) linear optical response as
depicted in Fig. \ref{fig:single-part-cond}, with $\sigma_{0}=e^{2}/4\hbar$. 

While the free--carrier response can be easily computed for the full four--band system, the same is not true when one wishes to obtain the excitonic response. The four possible combinations of valence and conduction bands would lead to a drastic increase in the computational complexity of both solving the Bethe--Salpeter equation and of obtaining the excitonic momentum matrix elements\cite{henriques_excitonic_2022,henriques_absorption_2022,quintela_tunable_2022}. Hence, it becomes crucial to reduce the system to a two--band problem when one wants to consider its excitonic properties. 
As seen in Fig. \ref{fig:single-part-cond}, and as expected from the large separation of the higher energy bands, we observe
minimal differences in the real part of the linear response between the four-- and two--band responses if $\hbar\omega\lesssim400\,\mathrm{meV}$,
consistent with previous results \cite{henriques_absorption_2022,quintela_theoretical_2022}. Hence, we will consider only the two $\eta=1$ bands from this point onward. This will, however, introduce issues when the imaginary part of $\sigma$ is relevant (i.e., when considering the magnitude of the various components of the conductivity tensor). Additionally, as only the two $\eta=1$ bands have an effect on the real part of the conductivity up to $\hbar\omega\approx E_{g}^{(12)}$ (for $V=55\,\mathrm{meV}$, up to $\hbar\omega\approx 400\,\mathrm{meV}$), one can expect minimal effects on the real part of the SHG nonlinear conductivity up to $\hbar\omega\approx E_{g}^{(12)}/2$. The imaginary part will, however, be significantly different from the four--band calculation much earlier due to both the longer range of the peaks in the imaginary part and the presence of resonances at $2\hbar \omega$. 

\section{Bethe--Salpeter Equation\label{sec:BSE}}

As discussed in the preceding section, we will focus on the dominant response
from the two bands closest to the bandgap $c_1,\, v_1$, which we denote $c,v$. Hence, we are only interested
in the eigenvalues $E_{c/v}$ and their respective eigenvectors, $\left|c/v\right\rangle $. Additionally, we will ignore contributions to eigenvectors from the skew coupling parameter, as this will allow us to transform the Bethe--Salpeter equation (BSE) from a 2D integral equation into a 1D integral equation. 

\begin{figure}
	\centering \hspace{0.3cm}\includegraphics[scale=0.72]{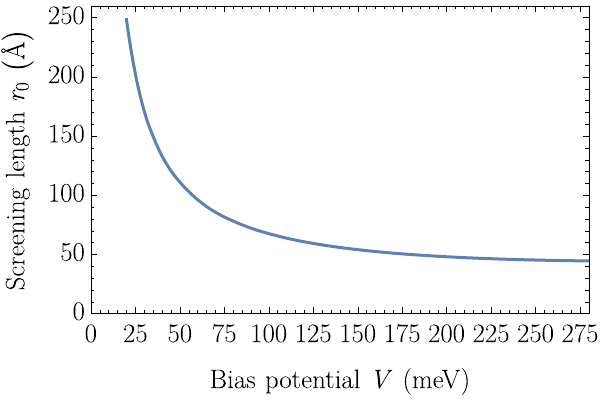}
	
	\caption{Evolution of the effective screening length $r_0$ as a function of external bias $V$. }\label{fig:r0_evolution}
\end{figure}

Before discussing the excitonic conductivity, we must first compute
the excitonic states. To compute the excitonic wave functions and binding energies, we will solve the BSE \cite{Glinskii1987,pedersen_intraband_2015,taghizadeh_nonlinear_2019,cao_unifying_2018,radha_optical_2021},
given in momentum space by 
\begin{equation}
\begin{aligned}E_n\psi_{cv\mathbf{k}}^{\left(n\right)} & =E_{cv\mathbf{k}}\psi_{cv\mathbf{k}}^{\left(n\right)}+\\
 & +\sum_{\mathbf{q}}V\left(\left|\mathbf{k}-\mathbf{q}\right|\right)\left\langle c_{\mathbf{k}}\middle|c_{\mathbf{q}}\right\rangle \left\langle v_{\mathbf{q}}\middle|v_{\mathbf{k}}\right\rangle \psi_{cv\mathbf{q}}^{\left(n\right)}
\end{aligned}
,\label{eq:bse}
\end{equation}
where $E_n$ is the exciton energy of state $n$, $V\left(k\right)$ is the
attractive electrostatic potential coupling the electron and the hole,
and $\psi_{cv\mathbf{k}}^{\left(n\right)}$ is the wave function of the
exciton. For our system, we consider $V\left(k\right)$ to
be the Rytova--Keldysh potential\cite{rytova_screened_1967,keldysh_coulomb_1979},
given in momentum space by 
\begin{equation}
V\left(k\right)=-2\pi\hbar c\alpha\frac{1}{k\left(\epsilon+r_{0}k\right)},\label{eq:rytova-keldysh}
\end{equation}
with $\alpha$ the fine--structure constant, $\epsilon$ the mean
dielectric constant of the media surrounding the bilayer, and $r_{0}$
as an in--plane screening length\cite{li_excitons_2019} related
to the polarizability of the material and usually obtained from DFT
calculations\cite{tian_electronic_2020}. The effective screening
length can be taken as a first approximation from the dipolar transition
amplitudes and is given by\cite{li_excitons_2019} 
\begin{align}
	r_{0}&=\frac{\hbar^{3}c\alpha}{\pi m_{0}^{2}}\int\frac{\left|\left\langle c_{\mathbf{k}}\left|P_{cv\mathbf{k}}^{x}\right|v_{\mathbf{k}}\right\rangle \right|^{2}}{E_{cv\mathbf{k}}^{3}}d^{2}\mathbf{k}.\label{eq:screen}
\end{align}
This effective screening length is very sensitive to the external bias, falling quickly as $V$ increases, as can be seen in Fig. \ref{fig:r0_evolution}.
For the considered bias potential, $r_{0}\approx103\,\text{\AA}$. Similarly to previous studies\cite{henriques_absorption_2022}, the dielectric constant is set at $\epsilon=6.9$ which corresponds to the the case of BBG encapsulated in hBN at the zero--frequency limit\cite{Laturia2018}. 

We then consider the excitonic wave function to have a well--defined
angular momentum $\ell_n$, writing it as 
\begin{align}
	\psi_{cv\mathbf{k}}^{(n)}=f_{cvk}^{(n)}e^{i\ell_{n}\theta_{k}}.
\end{align}
As the bands structure is isotropic, we replace $E_{cv\mathbf{k}}\rightarrow E_{cvk}$, allowing us to rewrite
the Bethe--Salpeter equation converting the sum into an integral
as 
\begin{widetext}
\begin{align}
E_n f_{cvk}^{(n)} & =E_{cvk}f_{cvk}^{(n)}+\frac{1}{4\pi^{2}}\sum_{\ell=-2}^{2}\int_0^{+\infty}\int_0^{2\pi} V\left(\left|\mathbf{k}-\mathbf{q}\right|\right)\mathcal{A}_{\ell}\left(k,q\right)e^{i\ell\tau\varphi}f_{cvq}^{(n)}e^{i\ell_{n}\varphi}d\varphi\,qdq,\label{eq:bse_full-1}
\end{align}
\end{widetext}
where $\varphi=\theta_{q}-\theta_{k}$. The radial component of the form factor can then be
written by analyzing the expansion of $\left\langle c_{\mathbf{k}}\middle|c_{\mathbf{q}}\right\rangle \left\langle v_{\mathbf{q}}\middle|v_{\mathbf{k}}\right\rangle $
as in terms of the $\ell$ factors in the complex exponential of the
form $e^{i\ell\tau\varphi}$ in Eq. (\ref{eq:bse_full-1}).
This then fixes the ranges on the sum over $\ell$ present in Eq.
(\ref{eq:bse_full-1}). For a monolayer system, the form factor reads,
in a somewhat abusive notation, 
\begin{align}
\mathcal{A}_{\ell}\left(k,q\right)e^{i\ell\tau\varphi} & =\left\langle c_{\mathbf{k}}\middle|c_{\mathbf{q}}\right\rangle \left\langle v_{\mathbf{q}}\middle|v_{\mathbf{k}}\right\rangle ,\label{eq:form-factor_monolayer}
\end{align}
identical to what is present in Eq. (\ref{eq:bse}). However, when one wishes
to consider a bilayer system where there is a distinction between
interlayer and intralayer phenomena, one must compute the form factor
more carefully. Recalling the eigenvectors from Eq. (\ref{eq:eigenvectors_simp}),
separated into top- and bottom--layer components as 
\begin{align}
\left|\lambda\right\rangle  & =\left[\begin{array}{c}
\left|\lambda^{t\;}\right\rangle \\
\left|\lambda^{b}\right\rangle 
\end{array}\right],
\end{align}
we explicitly introduce a distinction between intralayer and interlayer
interactions by writing the form factor as
\begin{align}
 & \mathcal{A}_{\ell}\left(k,q\right)e^{i\ell\tau\varphi}=\left\langle c_{\mathbf{k}}^{t}\middle|c_{\mathbf{q}}^{t}\right\rangle \left\langle v_{\mathbf{q}}^{t}\middle|v_{\mathbf{k}}^{t}\right\rangle +\left\langle c_{\mathbf{k}}^{b}\middle|c_{\mathbf{q}}^{b}\right\rangle \left\langle v_{\mathbf{q}}^{b}\middle|v_{\mathbf{k}}^{b}\right\rangle \nonumber \\
 & +e^{-d\left|\mathbf{k}-\mathbf{q}\right|}\left[\left\langle c_{\mathbf{k}}^{t}\middle|c_{\mathbf{q}}^{t}\right\rangle \left\langle v_{\mathbf{q}}^{b}\middle|v_{\mathbf{k}}^{b}\right\rangle +\left\langle c_{\mathbf{k}}^{b}\middle|c_{\mathbf{q}}^{b}\right\rangle \left\langle v_{\mathbf{q}}^{t}\middle|v_{\mathbf{k}}^{t}\right\rangle \right],\label{eq:form-factor_bilayer}
\end{align}
where $d\approx3.46\,\text{\AA}$ is the interlayer separation
between the two graphene layers\cite{doi:10.1021/ja994457o,Alam_2011,Butz2014,PhysRevB.104.L180202}. The extra $e^{-d\left|\mathbf{k}-\mathbf{q}\right|}$ factor present in Eq. (\ref{eq:form-factor_bilayer}) in fact originates from the Rytova--Keldysh potential, as written in Eq. (\ref{eq:rytova-keldysh}), corresponding to the charges being separated in the $z$-direction. It is, however, simpler to write this factor in the form factor as to explicitly couple only eigenvector components from opposing layers. 
As required, when the interlayer separation vanishes there will no longer be a vertical separation of the charges and the form factor exactly matches Eq. (\ref{eq:form-factor_monolayer}). 

Finally, Eq. (\ref{eq:bse_full-1}) is solved numerically via a simple
numerical quadrature in a tangent grid $q=\tan\left(x\frac{\pi}{2}\right)$,
with $1500$ points $x\in\left[0,1\right]$, following the procedure
already outlined several times in the literature, namely in Refs.
\cite{chao_analytical_1991,parfitt_two-dimensional_2002,henriques_absorption_2022,quintela_theoretical_2022}. Additionally, due to the introduction of the $e^{-d\left|\mathbf{k}-\mathbf{q}\right|}$ in Eq. (\ref{eq:form-factor_bilayer}), the angular integral in Eq. (\ref{eq:bse_full-1}) is done numerically in an uniform grid with $1500$ points $\varphi\in \left[0,2\pi\right]$.

\begin{figure}
	\centering\includegraphics[scale=0.71]{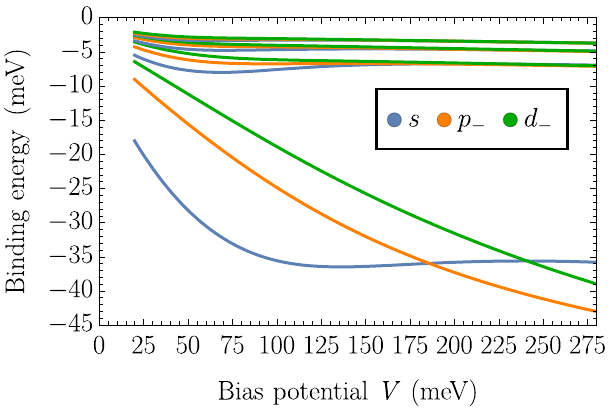}
	\caption{Evolution of the binding energies of the first four $s$, $p_-$, and $d_-$ -series excitons as a function of external bias $V$.}\label{fig:binding_evolution}
\end{figure}

When discussing excitonic states, we will use a similar nomenclature
as in the 2D Hydrogen atom\cite{PhysRevA.43.1186} to distinguish the different angular
momentum states (\emph{i.e.}, $s$, $p_{\pm}$, $d_{\pm}$ states). In Fig. \ref{fig:binding_evolution}, we plot the binding energy of the first four states of the $s$, $p_-$ and $d_-$ series. As the bias potential $V$ increases, the energy of the first $s$ and $p_-$ states approach each other, eventually crossing in the same way as presented in current \emph{ab-initio} computations\cite{sauer_exciton_2022}. In the model considered in this paper, this crossing occurs for a bias potential $V_{s-p_-}\approx180\,\mathrm{meV}$ and will lead to a swap of the first two peaks of the linear conductivity, as seen in Ref. \cite{sauer_exciton_2022}. The second crossing, involving $s$ and $d_-$ states, will occur at $V_{s-d_-}\approx245\,\mathrm{meV}$, and will lead to the dominant resonance for the linear OOP response being at a lower energy than the $1s$ resonance of the linear IP response. 

\section{Excitonic Optical Response\label{sec:optical}}

Having discussed the method for solving the BSE in a BBG system,
we now proceed to the study of the excitonic optical conductivity, writing the linear conductivity as\cite{pedersen_intraband_2015,taghizadeh_nonlinear_2019}
\begin{equation}
\frac{\sigma_{\alpha\beta}\left(\omega\right)}{\sigma_{0}}=\frac{-i\hbar^{2}}{2\pi^{3}m_0^{2}}\sum_{n}\left[\frac{E_{n}X_{0n}^{\alpha}X_{n0}^{\beta}}{E_{n}-\hbar\omega}-\left(\omega\rightarrow-\omega\right)^{*}\right],\label{eq:excitonic_first-order_simp1}
\end{equation}
where $\sigma_{0}=\frac{e^2}{4\hbar}$ is the conductivity of monolayer graphene. The excitonic matrix elements are
defined as\cite{pedersen_intraband_2015,taghizadeh_nonlinear_2019,PhysRevB.107.235416}
\begin{equation}
	X_{0n}^{\alpha}=i\int\,\psi_{cv\mathbf{k}}^{\left(n\right)}\frac{P_{vc\mathbf{k}}^{\alpha}}{E_{cv\mathbf{k}}}d^{2}\mathbf{k}.\label{eq:defs_X}
\end{equation}
For the nonlinear conductivity, we define $\sigma_{2}=\frac{e^{3}a}{4E_{g}\hbar}$ and write the SHG $\left(\omega_p = \omega_q \right)$ nonlinear conductivity\cite{pedersen_intraband_2015,taghizadeh_nonlinear_2019}
as
\begin{widetext}
	\begin{equation}
		\frac{\sigma_{\alpha\beta\gamma}^{\mathrm{SHG}}\left(\omega\right)}{\sigma_{2}}=\frac{-iE_{g}\hbar^{2}}{2a\pi^{3}m_0^{2}}\sum_{n,m}\left[\frac{E_{n}X_{0n}^{\alpha}Q_{nm}^{\beta}X_{m0}^{\gamma}}{\left(E_{n}-2\hbar\omega\right)\left(E_{m}-\hbar\omega\right)}-\frac{E_{n}X_{n0}^{\alpha}Q_{mn}^{\beta}X_{0m}^{\gamma}}{\left(E_{n}+2\hbar\omega\right)\left(E_{m}+\hbar\omega\right)}-\frac{\left(E_{n}-E_{m}\right)X_{0n}^{\alpha}Q_{nm}^{\beta}X_{m0}^{\gamma}}{\left(E_{n}+\hbar\omega\right)\left(E_{m}-\hbar\omega\right)}\right],\label{eq:excitonic_second-order_simp1-1}
	\end{equation}
\end{widetext}
where $E_g$ is the bandgap of the
material. As we are only considering effects from the two $\eta=1$ bands, we will take the bandgap $E_g$ as $E_{g}^{(11)}$, as discussed earlier. The two--state excitonic matrix elements are defined as\cite{pedersen_intraband_2015,taghizadeh_nonlinear_2019,PhysRevB.107.235416}
\begin{equation}
	Q_{nm}^{\alpha}=i\int\,\psi_{cv\mathbf{k}}^{\left(n\right)*}\left[\psi_{cv\mathbf{k}}^{\left(m\right)}\right]_{;k_{\alpha}}d^{2}\mathbf{k},\label{eq:defs_Q}
\end{equation}
where $\left[\psi_{cv\mathbf{k}}^{\left(m\right)}\right]_{;k_{\alpha}}$
is the generalized derivative\cite{aversa_nonlinear_1995} in the
$\alpha$--direction of the exciton wave function for the state
$m$ given in terms of the Berry connection $\Omega_{ij\mathbf{k}}^{\alpha}$,
defined as\cite{pedersen_intraband_2015}
\begin{equation}
	\left[\psi_{cv\mathbf{k}}^{\left(m\right)}\right]_{;k_{\alpha}}=\frac{\partial\psi_{cv\mathbf{k}}^{\left(m\right)}}{\partial k_{\alpha}}-i\left(\Omega_{cc\mathbf{k}}^{\alpha}-\Omega_{vv\mathbf{k}}^{\alpha}\right)\psi_{cv\mathbf{k}}^{\left(m\right)}.\label{eq:generalized_deriv-exciton}
\end{equation}
As the excitonic wave function will be independent of $k_{z}$, the
$\frac{\partial}{\partial k_{z}}\psi_{cv\mathbf{k}}^{\left(m\right)}$
term is dropped, meaning that $Q_{nm}^{z}$ reads\cite{PhysRevB.107.235416}
\begin{align}
	Q_{nm}^{z}&=\int\,\psi_{cv\mathbf{k}}^{\left(n\right)*}\left(\Omega_{cc\mathbf{k}}^{z}-\Omega_{vv\mathbf{k}}^{z}\right)\psi_{cv\mathbf{k}}^{\left(m\right)}d^{2}\mathbf{k}.\label{eq:Q_OOP}
\end{align}

\subsection{In--Plane Linear Response}

\begin{figure*}
	\begin{minipage}[c]{0.7\textwidth}
		\hspace{-0.923cm}\includegraphics[scale=0.9431]{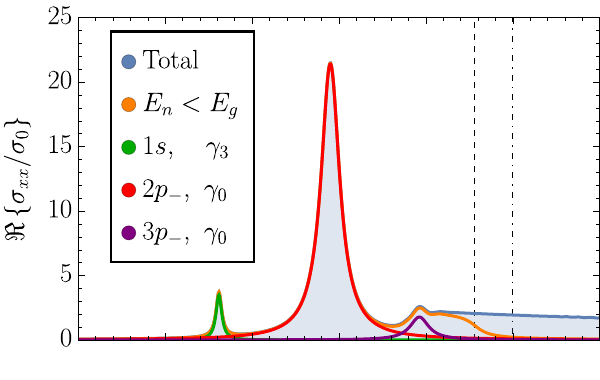}
		
		\vspace{-0.67cm}\hspace{-0.645cm}\includegraphics[scale=0.98]{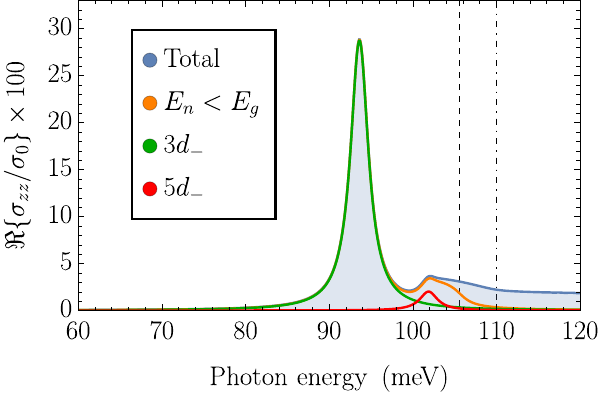}
	\end{minipage}%
	\begin{minipage}[c]{0.3\textwidth}
		\vspace{-1cm}\includegraphics[scale=0.7]{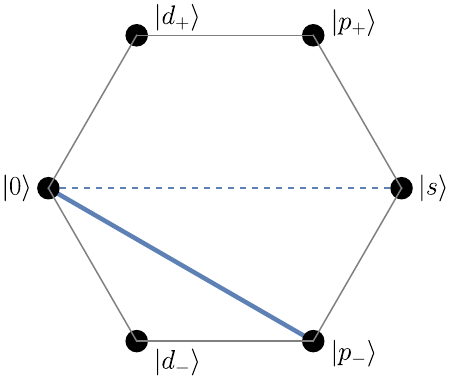}	
		
		\vspace{0.7cm}\includegraphics[scale=0.7]{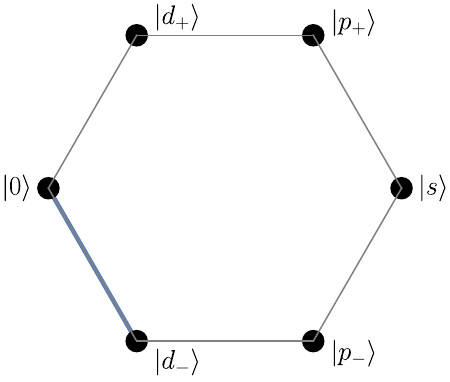}
	\end{minipage}
	\centering
	\caption{(Left) Real part of the linear IP (top) and OOP (bottom) optical response for $\epsilon_{\mathrm{medium}}=6.9$, with $\hbar\Gamma_{p_{-}}=\hbar\Gamma_{d_{-}}=1.3\,\mathrm{meV}$ and $\hbar\Gamma_{s}=0.4\,\mathrm{meV}$.
		Orange curve corresponds to the excitonic bound states, while blue line also includes continuum states. Vertical dashed line represents the bandgap $E_{g}^{(11)}$ of the system, dot-dashed line
		represents $2V=110\,\mathrm{meV}$. In the IP legend, $\gamma_{3}$ and $\gamma_{0}$ identify which matrix elements allow for the transition in question. 
		(Right) Diagram of dominant excitonic selection rules in the $\tau=1$ valley for linear IP (top) and OOP (bottom) optical response. Transitions allowed with/without skew coupling are shown in dashed/solid lines. 
		\label{fig:linear-excitonic-both}}
\end{figure*}	

Recalling Eq. (\ref{eq:defs_X}), the excitonic selection rules for IP linear response are obtained directly by considering the angular integral 
\begin{align}
\int_{0}^{2\pi}e^{i\ell_{n}\theta}P_{vc\mathbf{k}}^{x}d\theta,\label{eq:P_angular}
\end{align}
where $P_{vc\mathbf{k}}^{x}$ is as defined in Eq. (\ref{eq:expansion_P}).
Direct inspection of this integral tells us that the terms proportional to $\gamma_{0}$ will lead to contributions from states with $\ell_{n}=\pm\tau$,
while the presence of skew coupling $\gamma_{3}$ allows transitions to states where $\ell_{n}=\pm2\tau$. This means that the $X_{0n}^x$ matrix element can be written, in a somewhat abusive but concise form, as 
\begin{align}
	X_{0n}^x &=X_{\ell_n+\tau=0}^x+X_{\ell_n-\tau=0}^x\nonumber\\
	&\quad-i\frac{\gamma_{3}}{\gamma_{0}}\left[X_{\ell_n+2\tau=0}^x-X_{\ell_n-2\tau=0}^x\right],\label{eq:simp_X}
\end{align}
where the new indices restrict each term to the Kronecker $\delta$'s resulting from the different angular integrals.
Taking $\left|X_{0n}^x\right|^2$ as in Eq. (\ref{eq:excitonic_first-order_simp1}), the presence of the various Kronecker $\delta$'s originating from each of the angular integrals means that non--vanishing contributions from $\gamma_{3}$ to the linear response must be proportional to $\left|\gamma_3\right|^2$. Therefore, the inclusion of an $i$ factor in the Hamiltonian will not alter the results when compared to previous studies\cite{henriques_absorption_2022}. It will, however, affect the nonlinear response as it will be linear in $\gamma_{3}$ (in this simple approximation) and, therefore, the $i$ factor proves to be important.
\begin{figure}
	\includegraphics[scale=0.7]{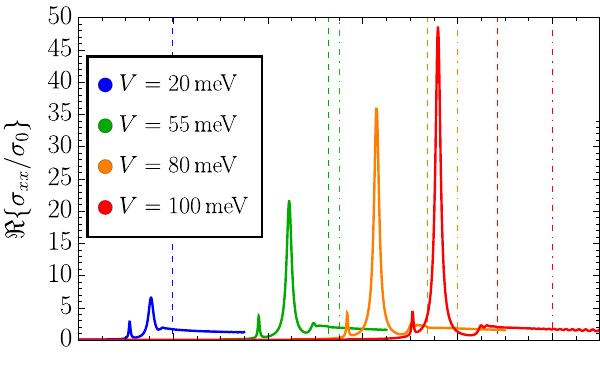}

	\vspace{-0.97cm}\hspace{0.7cm}\includegraphics[scale=0.7]{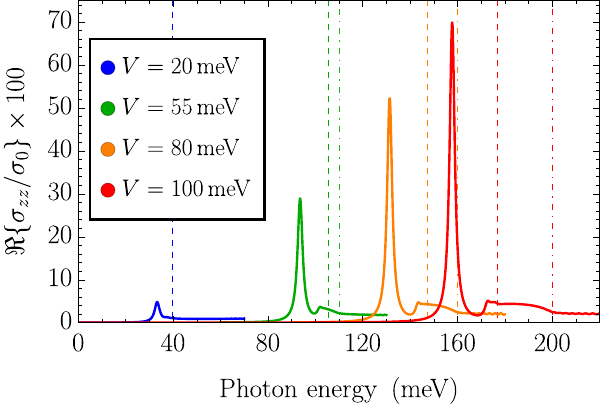}	
		
	\centering
	\caption{Real part of the linear IP (top) and OOP (bottom) optical response for $\epsilon=6.9$, with $\hbar\Gamma_{p_{-}}=\hbar\Gamma_{d_{-}}=1.3\,\mathrm{meV}$ and $\hbar\Gamma_{s}=0.4\,\mathrm{meV}$, for values of the bias potential $V$ between $20$ and $100\,\mathrm{meV}$. 
	Vertical dashed lines represent the bandgap $E_{g}^{(11)}$ of the system for each value of the bias potential $V$, dot-dashed lines represent the gap at $k=0$, equal to $2V$.
	\label{fig:linear-excitonic-both_V}}
\end{figure}	

Following from the results of Ref. \cite{ju_tunable_2017}, we will assume the broadening to be dependent on the excitonic state $n$. Specifically, we set the broadening as dependent on the angular momentum series, given by $\hbar\Gamma_{p_{-}}=1.3\,\mathrm{meV}$ and $\hbar\Gamma_{s}=0.4\,\mathrm{meV}$. With the broadening set, we plot the IP linear response for $V=55\,\mathrm{meV}$ in the top panel of Fig. \ref{fig:linear-excitonic-both}. Varying the bias potential $V$, we can observe the tunability of the IP linear response in the top panel of Fig. \ref{fig:linear-excitonic-both_V}. Similarly to what was observed in Refs. \cite{henriques_absorption_2022,sauer_exciton_2022}, the dominant peak for the IP linear response quickly increases as the bias potential increases. Additionally, the location of the peak corresponding to the $1s$ resonance approaches that of the dominant $2p_-$ resonance, as indicated by the evolution of the binding energies in Fig. \ref{fig:binding_evolution}.

\subsection{In--Plane Nonlinear Response}

Now for the nonlinear optical response, we recall Eqs. (\ref{eq:defs_Q}-\ref{eq:generalized_deriv-exciton}) and write the integrand  present in the two--state momentum matrix element as
\begin{align}
	&\psi_{cv\mathbf{k}}^{\left(n\right)*}\left[\frac{\partial\psi_{cv\mathbf{k}}^{\left(m\right)}}{\partial k_{x}}-i\left(\Omega_{cc\mathbf{k}}^{x}-\Omega_{vv\mathbf{k}}^{x}\right)\psi_{cv\mathbf{k}}^{\left(m\right)}\right].\label{eq:Q_x_matrix}
\end{align}
Focusing on the angular integration of this expression, and recalling the definition of 
the Berry connection present in Eq. (\ref{eq:omega_cc_simp}), one obtains that only transitions with $\ell_{m}-\ell_{n}=\pm1$ are allowed. In a similar notation to Eq. (\ref{eq:simp_X}), this means that the Berry connection up to zeroth order in $\frac{\gamma_{3}}{\gamma_{0}}$ can be written as 
\begin{equation}
	Q_{nm}^{x}=Q_{\left|\ell_{m}-\ell_{n}\right|=1}^x\label{eq:simp_Q}.
\end{equation}
Multiplication of the angular momentum Kronecker $\delta$'s from Eqs. (\ref{eq:simp_X}-\ref{eq:simp_Q}) leads to four non--zero matrix elements which will be linear in $\frac{\gamma_{3}}{\gamma_{0}}$. We denote these non--zero matrix elements by the oscillator strength defined, for compactness, as 
\begin{align}
	\sigma_{\ell_{n};\ell_{m}}&\equiv X_{\ell_n}^x Q_{\ell_n,\ell_m}^x X_{\ell_m}^x.
\end{align}
Explicitly, the transitions allowed by Eqs. (\ref{eq:P_angular}-\ref{eq:Q_x_matrix}) will be 
\begin{align}
	&\sigma_{s;p_-}&&\hspace{-0.8cm}\sigma_{p_-;s}\nonumber\\
	&\sigma_{f_-;g_-}&&\hspace{-0.8cm}\sigma_{g_-;f_-},
\end{align}
where the left column corresponds to those where $\ell_{m}-\ell_{n}=-1$ and the right column to those where $\ell_{m}-\ell_{n}=1$. 

Careful inspection of the matrix elements tells us that $\sigma_{s;p_-}$ and $\sigma_{p_-;s}$ will be the dominant contributions to the SHG nonlinear response up to linear order in $\frac{\gamma_{3}}{\gamma_{0}}$. The real part of the SHG nonlinear response is then plotted in the top panel of Fig. \ref{fig:nonlinear-excitonic-both}. Similarly to previous results\cite{PhysRevB.107.235416}, we see that the dominant contribution will originate from the excitonic states with the smallest angular momentum. 

\subsection{Out--of--Plane Linear Response}

Recalling Eq. (\ref{eq:defs_X}), the dipole transition amplitude for the excitonic state $n$ with angular momentum quantum number $\ell_{n}$ reads 
\begin{align}
X_{0n}^{z}
&=\int_{0}^{\infty}\frac{f_{cvk}^{(n)}}{E_{cvk}}kdk\int_{0}^{2\pi}e^{i\ell_{n}\theta}P_{vc\mathbf{k}}^{z}d\theta
\end{align}
which immediately leads to the optical selection rule $\ell_{n}=0$
for the linear response when one recalls the definition of $P_{vc\mathbf{k}}^{z}$ from Eq. (\ref{eq:expansion_P_z}). Inspecting this equation also tells us that the OOP linear optical response will have a quadratic dependence on the interlayer spacing, originating directly from 
$\left|\int_{0}^{2\pi}e^{i\ell_{n}\theta} P_{vc\mathbf{k}}^z \,d\theta\right|^2$.
Additionally, a more complex dependence on $d$ will be implicitly present in the excitonic wave function $f_{cvk}^{(n)}$ due to the exponential $e^{-d\left|\mathbf{k}-\mathbf{q}\right|}$ in the form factor present in Eq. (\ref{eq:form-factor_bilayer}). 

The real part of the OOP linear excitonic optical response is plotted in the bottom panel of Fig. \ref{fig:linear-excitonic-both}, together with the diagram of the allowed transitions in the $\tau=1$ valley. Although it is not labelled in the figure due to its much weaker amplitude, the $4d_-$ transition is still allowed. Contrary to what was computed for the OOP linear response of a buckled monolayer\cite{PhysRevB.107.235416}, we observe that the conductivity quickly stabilizes to a constant value for $\hbar\omega > E_{g}^{(11)}$, while in a buckled monolayer an apparently linear growth with $\hbar \omega$ was observed in the same regime. The observed behaviour also matches with the results for the free--carrier conductivity, present in the bottom panel of Fig. \ref{fig:single-part-cond}, in the region where the two--band approximation is valid. Additionally, in the bottom panel of Fig. \ref{fig:linear-excitonic-both_V}, we present the OOP optical response for various values of the bias potential $V$ between $20$ and $100\,\mathrm{meV}$. In this figure, we can see the much larger effect of the bias potential on the OOP optical response in comparison to the IP response. This is most noticeable between $V=20\,\mathrm{meV}$ and $55\,\mathrm{meV}$, where the maximum of the linear response for the IP and OOP linear response increases by factors of approximately $3\times$ and $5\times$, respectively. 

\subsection{Out--of--Plane Nonlinear Response}

\begin{figure*}
	\begin{minipage}[c]{0.7\textwidth}
		\hspace{-0.347cm}\includegraphics[scale=.986]{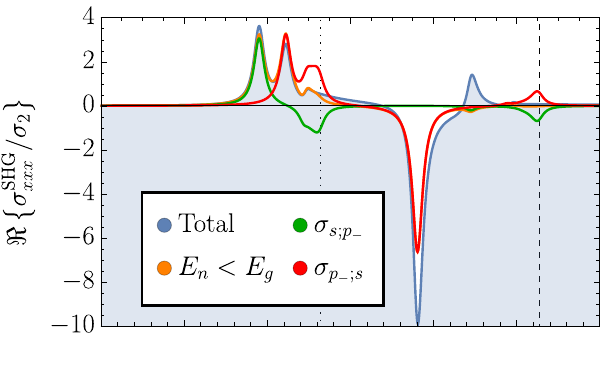}
		
		\vspace{-0.725cm}\hspace{-0.5cm}\includegraphics[scale=1]{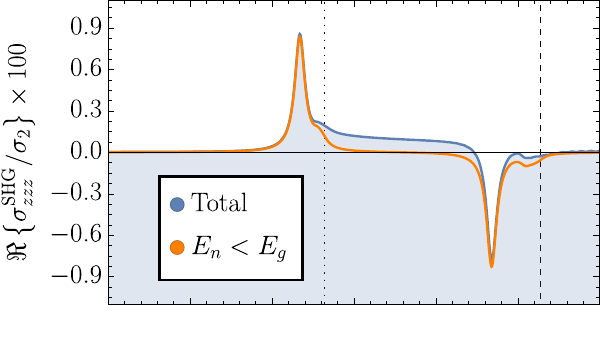}	
		
		\vspace{-0.71cm}\hspace{-0.077cm}\includegraphics[scale=1.0255]{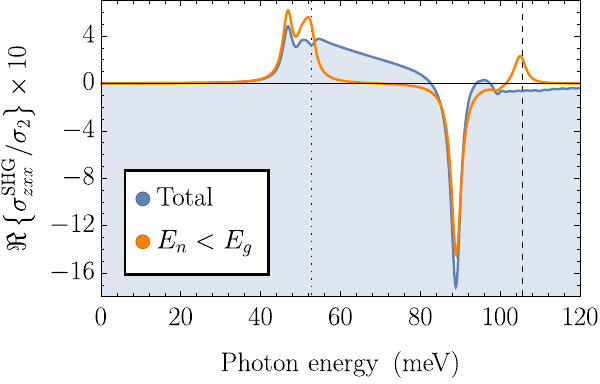}
	\end{minipage}%
	\begin{minipage}[c]{0.3\textwidth}
		\vspace{-1.3cm}\includegraphics[scale=0.7]{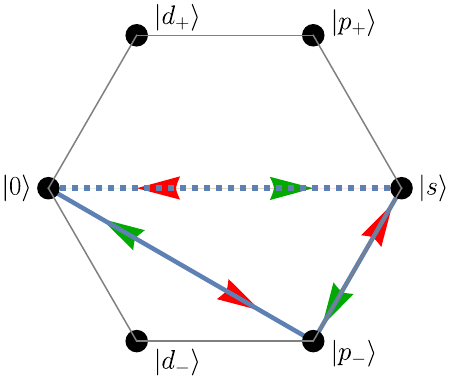}
		
		\vspace{0.7cm}\includegraphics[scale=0.7]{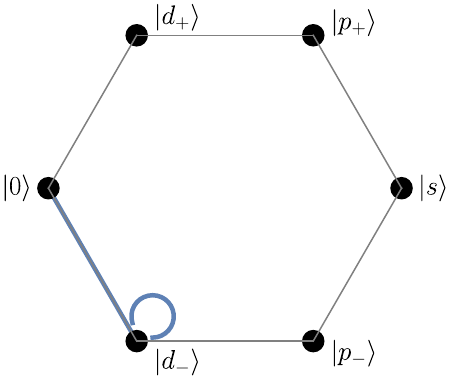}
		
		\vspace{0.7cm}\includegraphics[scale=0.7]{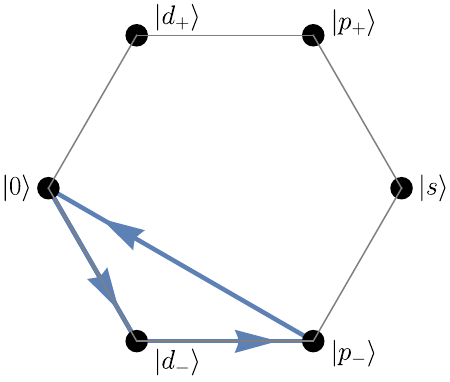}
	\end{minipage}
	\centering
	\caption{(Left) Real part of the SHG optical response of BBG with diagonal IP (top), diagonal OOP (middle) and non--diagonal OOP (bottom) conductivity for $\epsilon_{\mathrm{medium}}=6.9$ and $\hbar\Gamma=1.5\,\mathrm{meV}$. Orange curve corresponds to only excitonic bound states, while blue line also includes continuum states. Vertical (dotted) dashed black lines represent (half) the bandgap $E_{g}^{(11)}$ of the system. 
		(Right) Diagram of dominant excitonic selection rules in the $\tau=1$ valley for each component. Transitions allowed with/without skew coupling are shown in dashed/solid lines. Arrow direction and colour represent the specific resonance when multiple contributions are present. 
		\label{fig:nonlinear-excitonic-both}}
\end{figure*}

Focusing now on the OOP nonlinear regime, we recall the expression for $Q_{nm}^{z}$ from Eq. (\ref{eq:Q_OOP}) as
\begin{align}
	Q_{nm}^{z}=\int\psi_{cv\mathbf{k}}^{\left(n\right)*}\left(\Omega_{cc\mathbf{k}}^{z}-\Omega_{vv\mathbf{k}}^{z}\right)\psi_{cv\mathbf{k}}^{\left(m\right)}d^{2}\mathbf{k}.
\end{align}
With this expression in mind, careful analysis of Eq. (\ref{eq:berry_z})
leads immediately to the selection rules
\begin{align}
	\ell_{n}&=\ell_{m}.
\end{align}
This selection rule is rather unrestrictive, meaning that it will be $X_{0n}^z$ which will determine the allowed transitions. 

Focusing first on the diagonal $\sigma_{zzz}$ response, the selection
rules follow immediately from analysis of $X_{0n}^{z}$ and $Q_{nm}^{z}$ as
\begin{align}
	\ell_{n}&=\ell_{m}=0.
\end{align}
Recalling the pseudo--spin factor, this means that the transitions present will be associated with $d_-$--series states. 
Knowing the selection rules, we can compute the SHG conductivity, plotted in the middle panel of Fig. \ref{fig:nonlinear-excitonic-both}. The matrix elements $X_{0n}^z$ (around $1/5$ the magnitude of $X_{0n}^x$) and $Q_{nm}^z$ (around $1/4$ the magnitude of the analogous matrix element in $Q_{nm}^x$) are significantly smaller than those present in $\sigma_{xxx}^\mathrm{SHG}$. This,  in conjunction with the cubic dependence\cite{PhysRevB.107.235416} on the ratio $d/a\approx0.7$, means that the peak amplitude of $\sigma_{zzz}^{\mathrm{SHG}}$ is close to $800$ times smaller than that of $\sigma_{xxx}^\mathrm{SHG}$.
A big difference can also be observer in the qualitative behaviour of $\sigma_{zzz}^{\mathrm{SHG}}$ versus its analogous counterpart in a buckled monolayer\cite{PhysRevB.107.235416}: while in the buckled monolayer the $\sigma_{zzz}^{\mathrm{SHG}}$ response above the bandgap remains close to its maximum, in BBG we observe that it quickly tends to a much smaller value, similarly to what is observed in the $\sigma_{xxx}^{\mathrm{SHG}}$ response. 

Secondly, we focus our attention on the off-diagonal $\sigma_{zxx}$
response. While the presence of $X_{0n}^{z}$ means immediately that
$\ell_{n}=0$, we must now carefully analyze Eqs. (\ref{eq:expansion_P}-\ref{eq:P_angular}), as well as Eqs. (\ref{eq:omega_cc_simp}-\ref{eq:Q_x_matrix}).
From $Q_{nm}^{x}$, we immediately get that 
\[
\left|\ell_{m}-\ell_{n}\right|=1\Rightarrow\ell_{m}=\pm1,
\]
both allowed by the selection rules for $X_{m0}^{x}$. Hence, we are restricted to states
\begin{align*}
\ell_{n} & =0, & \ell_{m} & =\pm1.
\end{align*}
Again recalling the pseudo--spin factor, we can immediately associate the selection rules with transitions between $d_-$ states and both $p_-$ and $f_-$ states. Careful inspection of the oscillator strength for the two distinct selection rules reveals that, as one would intuitively expect from the lower angular momentum, resonances associated with the matrix elements $\sigma_{d_-;p_-}$ dominate. The off-diagonal $\sigma_{zxx}^{\mathrm{SHG}}$
SHG response is then plotted in the bottom panel of Fig. \ref{fig:nonlinear-excitonic-both}.  

\begin{figure}
	\hspace{-0.875cm}\includegraphics[scale=0.7172]{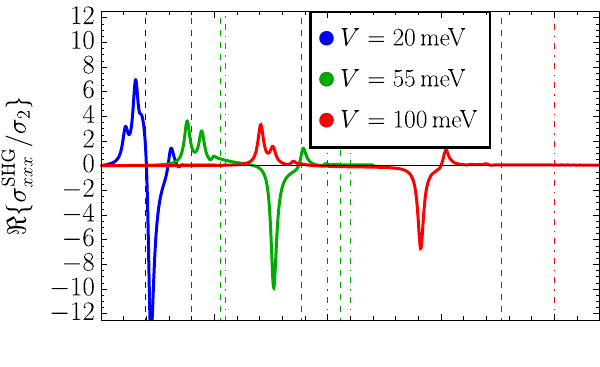}
	
	\vspace{-0.61cm}\hspace{-0.7cm}\includegraphics[scale=0.6998]{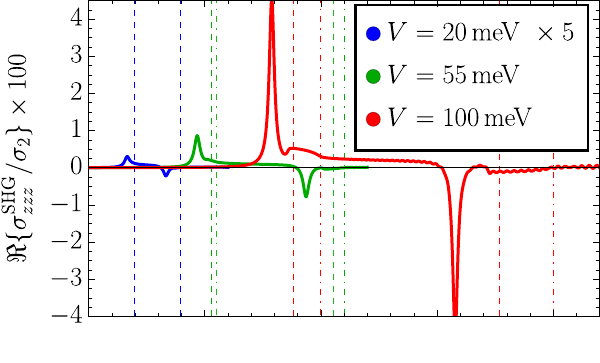}
	
	\vspace{-0.59cm}\hspace{-0.875cm}\includegraphics[scale=0.7172]{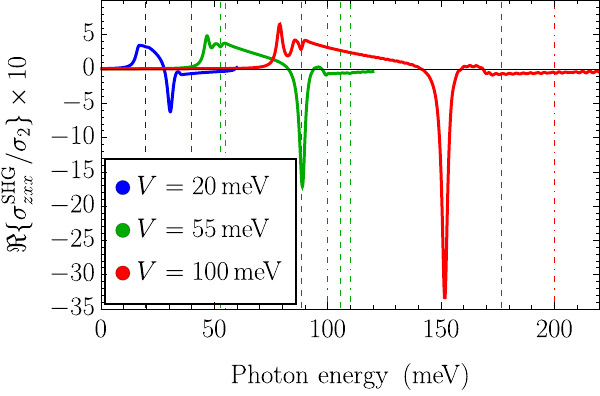}	
	
	\centering
	\caption{Real part of the SHG IP (top), diagonal OOP (middle) and nondiagonal OOP (bottom) optical response for $\epsilon=6.9$, with broadening $\hbar\Gamma=1.5\,\mathrm{meV}$, for values of the bias potential $V$ between $20$ and $100\,\mathrm{meV}$. 
	Vertical dashed lines represent the bandgap $E_{g}^{(11)}$ of the system for each value of the bias potential $V$, dot-dashed lines represent the gap at $k=0$, equal to $2V$. 
	Leftmost vertical lines correspond to response at $2\hbar\omega$, rightmost vertical lines correspond to response at $\hbar\omega$.\label{fig:SHG-excitonic-both_V}}
\end{figure}

Finally, in Fig. \ref{fig:SHG-excitonic-both_V}, we present the SHG IP (top), diagonal OOP (middle) and nondiagonal OOP (bottom) optical response for three different values of the bias potential $V$. In this figure, we keep a constant broadening $\hbar\Gamma=1.5\,\mathrm{meV}$ as to more easily observe the effects of tuning the bias potential. For the lowest value of the bias potential, $V=20\,\mathrm{meV}$, the lowest bandgap $E_{g}^{(11)}=39.8\,\mathrm{meV}$ is very close to $2V=40\,\mathrm{meV}$, we only keep the vertical lines associated with $E_{g}^{(11)}$ and $E_{g}^{(11)}/2$ for improved readability. Starting with the IP response $\sigma_{xxx}^{\mathrm{SHG}}$, we can observe that, at $V=20\,\mathrm{meV}$, the proximity between the multiple resonances leads to a sharp increase in intensity when compared with the response at $V=55\,\mathrm{meV}$ and $100\,\mathrm{meV}$. For the diagonal OOP response $\sigma_{zzz}^{\mathrm{SHG}}$, we observe that the magnitude of the excitonic response quickly grows with the bias potential $V$. This follows what is expected, as it is the OOP anisotropy due to the bias potential that leads to a non--zero OOP response. Lastly, the nondiagonal OOP response $\sigma_{zxx}^{\mathrm{SHG}}$ also presents an increasing magnitude with the bias potential $V$, as expected from the growing OOP anisotropy. The nature of the nondiagonal response, with its mixing of IP and OOP components, means that the response at low bias potential (see $V=20\,\mathrm{meV}$ regime) does not vanish as quickly as the $\sigma_{zzz}^{\mathrm{SHG}}$ component. Additionally, direct comparison between top and bottom panels of Fig. \ref{fig:SHG-excitonic-both_V} shows that, as bias potential increases, the magnitude of the nondiagonal OOP response approaches that of the IP response, with the maximum of $\sigma_{zxx}^{\mathrm{SHG}}$ approximately $1/2$ that of $\sigma_{xxx}^{\mathrm{SHG}}$ for $V=100\,\mathrm{meV}$. 

\section{Summary}

In this paper, we studied the excitonic linear and nonlinear optical response of Bernal stacked BBG as a function of the gate voltage, both for in--plane (IP) and out--of--plane (OOP) directions. Starting with the electronic structure of BBG, we consider the influence of gate voltage on the band structure of the system. Taking into account the three distinct bandgaps present, we introduce a two--band approximation that greatly simplifies the numerical complexity of the problem. Outlining the generic form of the momentum matrix elements and Berry connections, we compute the linear IP and OOP free--carrier optical response for both the two--band approximation and the full four--band Hamiltonian. The large separation of the higher energy bands (for $V=55\,\mathrm{meV}$, the lowest bandgap is $E_{g}^{(11)}=106\,\mathrm{meV}$ while the second lowest is $E_{g}^{(12)}=436\,\mathrm{meV}$) means that, for frequency  below $E_{g}^{(12)}$, the real part of the optical response is very accurately described by the two--band approximation. 

Under this two--band approximation, we then computed the excitonic states in the system by numerical diagonalization of the Bethe--Salpeter equation, studying the evolution of the excitonic binding energies as a function of external bias potential. 

Knowing the excitonic states, together with the momentum matrix elements and Berry connections, allowed us to then explicitly discuss the excitonic selection rules of the system for both the IP and OOP excitonic response. The inclusion of skew coupling in the momentum matrix elements proves fundamental in obtaining a non--zero IP nonlinear response, as well as introducing $s-$series resonances in the IP linear response. For the OOP linear response the OOP anisotropy proves to be sufficient for obtaining a non--zero linear and nonlinear response.

Both linear and SHG nonlinear response are very sensitive to the tuning of the bias voltage. Firstly, tuning the bias potential will directly affect the bandgap of the system, leading to a tuning of the excitonic binding energies and, therefore, the location of the resonant peaks in the optical response, as well as on the interband momentum matrix elements. Additionally, as the OOP anisotropy is controlled by the bias potential, its effects on the OOP response are not as simple as those on the IP response. 

\section*{Acknowledgements}

M.F.C.M.Q. acknowledges the International Iberian Nanotechnology Laboratory
(INL) and the Portuguese Foundation for Science and Technology (FCT)
for the Quantum Portugal Initiative (QPI) grant SFRH/BD/151114/2021.
N.M.R.P. acknowledges support by the Portuguese Foundation for Science and Technology (FCT) in the framework of the Strategic Funding UIDB/04650/2020, COMPETE 2020, PORTUGAL 2020, FEDER, and  FCT through projects PTDC/FIS-MAC/2045/2021, EXPL/FIS-MAC/0953/2021. N.M.R.P. also acknowledges the Independent Research Fund Denmark (grant no. 2032-00045B) and the Danish National Research Foundation (Project No. DNRF165).

\bibliographystyle{ieeetr}
\bibliography{nlo_bilayer}
 
\end{document}